\documentclass[preprint,12pt,numafflabel]{elsarticle}
\usepackage{amsmath, amssymb, amscd, amsthm, amsfonts}
\usepackage[title]{appendix}
\usepackage{graphicx}
\usepackage{hyperref}
\usepackage{color}
\usepackage{subcaption}
\usepackage{float}
\usepackage[section]{placeins}
\usepackage{indentfirst}
\usepackage{comment}

\usepackage{diffcoeff}

\begin{document}
\begin{frontmatter}


\title{Performance Studies of the Mu2e Cosmic Ray Veto Detector}
\author[1]{Simon Corrodi}
\author[2]{Mackenzie Devilbiss}
\author[3]{E. Craig Dukes}
\author[3]{Ralf Ehrlich}
\author[3]{R. Craig Group}
\author[3]{Tyler Horoho}
\author[1]{Yuri Oksuzian}
\author[4]{Paul Rubinov}
\author[3]{Matthew Solt}
\author[1]{Yongyi Wu\corref{cor1}}
\ead{ywu7@bu.edu}
\affiliation[1]{Argonne National Laboratory, Lemont, IL 60439, USA}
\affiliation[2]{University of Michigan, Ann Arbor, MI 48109, USA}
\affiliation[3]{University of Virginia, Charlottesville, VA 22904, USA}
\affiliation[4]{Fermi National Accelerator Laboratory, Batavia, IL 60510, USA}
\cortext[cor1]{Corresponding author, now at Boston University}
\date{\today}

\begin{abstract}
The cosmic ray veto (CRV) detector of the Mu2e experiment consists of four layers 
of plastic scintillation counters that surround the detector solenoid. These counters 
are embedded with wavelength-shifting fibers and are read out by silicon 
photomultipliers (SiPMs). The performance of a subset of the CRV counters was studied in a 
cosmic-ray test stand. Using data taken over a two-year period, we report the single-layer 
muon detection efficiency and the rate at which the light yield degrades due to the aging of the plastic scintillation counters.  
\end{abstract}
\begin{keyword}
Cosmic Ray Veto \sep plastic scintillator \sep efficiency \sep aging \sep Mu2e
\end{keyword}
\end{frontmatter}

\newpage
\pagenumbering{arabic}
\tableofcontents
\newpage

\section{Introduction}\label{sec:into}
The Mu2e experiment intends to search for the conversion of a negatively charged muon 
into an electron in the presence of an aluminum nucleus: 
$\mu^- {\rm Al} \rightarrow e^- {\rm Al}$~\cite{Mu2e:2014fns}.  In the Standard Model 
this process is forbidden. In the extended Standard Model with non-zero neutrino masses, the ratio of conversion to capture is at an experimentally unobservable value of 
$R(\mu^-{\rm Al} \rightarrow e^-{\rm Al}) \simeq 10^{-52}$~\cite{Marciano:2008zz}. Hence, any observation of this process would be unambiguous 
evidence of new physics.

For Mu2e to achieve its goals, it is vital to keep backgrounds under control, if possible to less than one event 
over the duration of the experiment. 
A simulation effort has been made to estimate the various backgrounds: 
see Ref.\,\cite{SU2020} for details.  The backgrounds come in three types: 
(1) stopped-muon-induced backgrounds; (2) beam-related backgrounds;
and (3) time-dependent backgrounds.
\begin{figure}[htbp]
	\centerline{\includegraphics[width=6.0in]{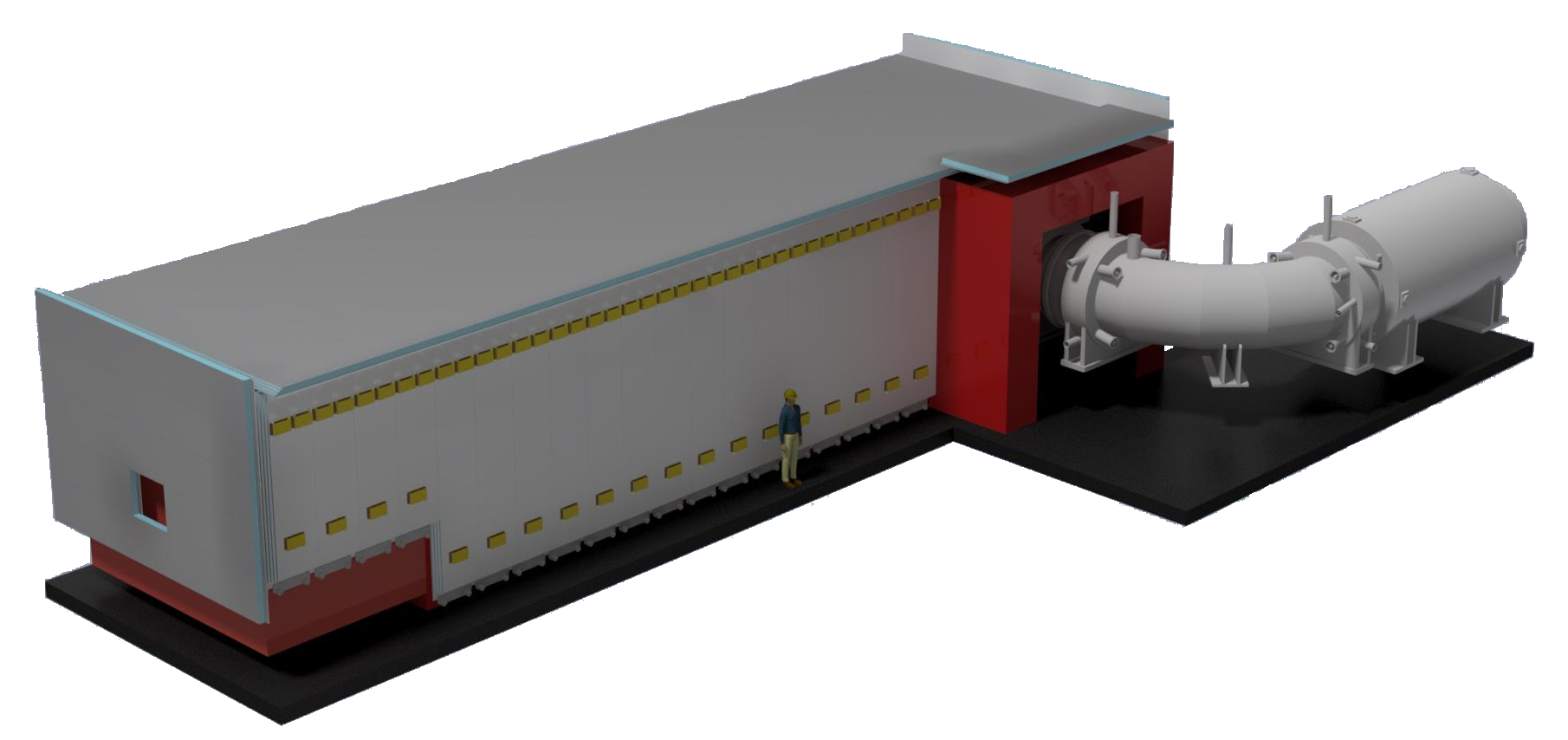}}
  \caption{
   The cosmic ray veto (gray) consists of 83 modules that are mounted on the top and sides of concrete 
   shielding blocks (red) that surround the Mu2e apparatus, which is enclosed within a solenoidal magnet 
   downstream of another S-shaped solenoidal magnet that is seen at the right.}
  \label{fig:crv}
\end{figure}

Cosmic rays, as a source of time-dependent backgrounds, can produce conversion-like electrons
from muon decays-in-flight, neutron interactions, and muons misidentified as 
conversion electrons.  Such events are expected to appear at a rate of roughly 
one per day and represent the largest background to the experiment.  To identify and veto
such events, a cosmic ray veto (CRV) detector has 
been fabricated that will surround much of the Mu2e apparatus (Fig.\,\ref{fig:crv}).  
The CRV is designed to veto cosmic-ray muons with a $99.99 \%$ efficiency, reducing the cosmic-ray background expectation to less than one event over the full experimental run. 

\section{The Cosmic Ray Veto Detector Design}\label{sec:crv}
The CRV consists of four layers of scintillator counters with embedded wavelength-shifting fibers, read out by silicon photomultipliers.
A conversion-like electron in the apparatus that comes in coincidence with a track stub found in the CRV is vetoed in the offline analysis.

The CRV scintillator counters are grouped in modules, each module (with one exception) consisting of 64 counters, 16 per layer, with aluminum sheets of 9.525\,mm (3/8") thickness placed between each layer.
The entire CRV consists of 83 total modules.
Each scintillator counter is $51.34{\times}19.80$\,mm$^2$ in cross section, with lengths ranging from 0.985\,m to 6.900\,m.\footnote{The counters have a polystyrene base doped with 1\% PPO, and 0.03\% POPOP.}
The counters have a thin TiO$_2$ reflective coating and two channels into which the wavelength-shifting (WLS) fibers are inserted~\footnote{Kuraray America, Inc.}.

\begin{figure}[htbp]
\centering
\includegraphics[scale=0.4]{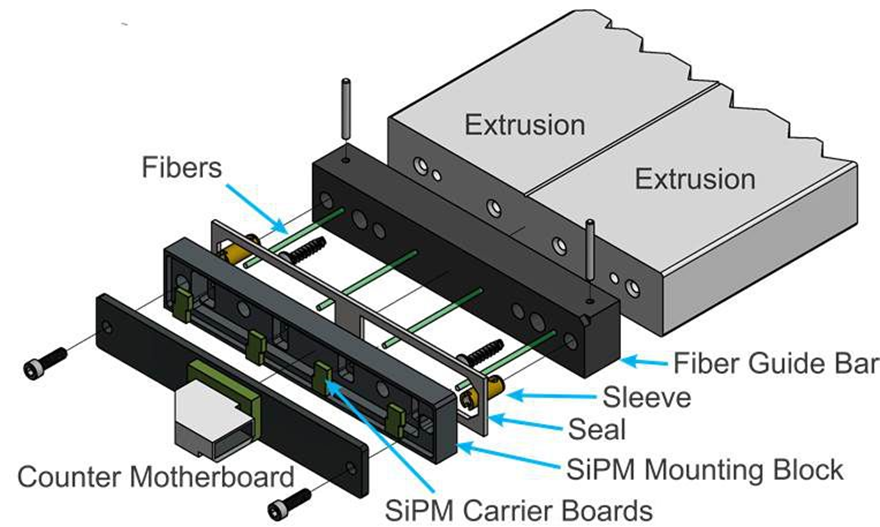}
\caption{Exploded view of the manifold at the end of a di-counter showing the fiber guide bar, SiPM mounting block, SiPM carrier boards, SiPMs, and counter motherboard. The flasher LEDs, thermometer, and pogo pins are not shown.\label{fig:dicounterreadout}}
\end{figure}

The fibers are read out by $2{\times}2$\,mm$^2$ silicon photomultipliers (SiPMs) located at each end, except in special cases where only one end is read out.\footnote{The SiPMs are Hamamatsu S13360-2050VE devices with a 50\,$\mu$m pixel size, the large size being chosen as a compromise between dynamic range and quantum efficiency.}
For readout purposes, two counters are glued together side-by-side to form a di-counter (Fig.~\ref{fig:dicounterreadout}).
An acetal fiber guide bar (FGB) is attached to each end of a di-counter, into which the fibers are glued, after which their ends are polished. (The fibers are not glued into the scintillator channels.)
Each SiPM is mounted on a tiny printed circuit board by the manufacturer~\footnote{Hamamatusu Corp.}(called SiPM carrier boards, see Fig.~\ref{fig:dicounterreadout})). 
The SiPMs are housed in small rectangular wells in a black anodized aluminum fixture called the SiPM mounting block, which is precisely registered to the FGB by a pair of sleeves.
A small circuit board, the counter motherboard (CMB), has spring-loaded ``pogo" pins that gently press the SiPMs against the fibers and provide electrical  contact to the SiPM carrier boards.
The CMB provides the SiPM bias voltage and sends the four SiPM output signals of each di-counter end to a readout board via an HDMI cable.
The CMB also has two flasher LEDs, one for 
each counter, and a single thermometer centered on the board.\footnote{The thermometer is a UMW DS18B20U digital temperature sensor with a range from -55$^{\circ}$C to 125$^{\circ}$C and an accuracy of $\pm 0.4^{\circ}$C.} 
This thermometer's temperature measurement is used as a proxy for the temperature of the four SiPMs serviced by the CMB.

The readout manifold attached to each di-counter end was designed to be as compact as possible to minimize
gaps in the coverage of the CRV that could allow cosmic-ray muons to leak through undetected.  Cooling the SiPMs to control their temperature,
and hence breakdown voltage, as well as to reduce their dark count rates, was
entertained, but rejected due to the strict gap requirements (needed to keep the
detection efficiency for cosmic-ray muons over 99.99\%) as well as cost considerations.
Rather, it was decided to carefully monitor the SiPM temperatures to allow for 
temperature corrections to be made to correctly determine the 
photoelectron (PE) yields.

All of the 83 modules that comprise the CRV have been fabricated and are
at Fermilab in the Wideband detector hall awaiting installation~\cite{Boi:2021yxw}.
The tests reported here come from several modules situated in a cosmic-ray
test stand.

\section{Readout of the Cosmic Ray Veto}\label{sec:readout}

Below is an abbreviated description of the CRV readout system.
For a more detailed description, see~\cite{crvsipm}.
The main component of the CRV readout system is the front-end board (FEB).  
It provides bias to the SiPMs and amplifies, shapes, digitizes, zero-suppresses, and stores the signals coming from the SiPMs.
Each FEB serves 64 SiPMs via 16 HDMI cables: two FEBs are required to read out each end of a CRV module.
In the experiment, the FEBs are installed within steel magnetic enclosures to shield their inductors and transformers from the large ambient field of the detector solenoid.
The FEBs are powered and read out using CAT6 Ethernet cables connected to a readout controller (ROC), 
which also collects data from multiple FEBs for transmission to the Data Transfer Controllers (DTCs) 
that serve as the interface to the data acquisition (DAQ) computers.
However, in this work, the ROCs only supplied power, timing, and triggers, while data were read directly from the FEBs to a DAQ PC, bypassing the DTCs.
The DAQ PC also set the amplifier gain, SiPM biases, and monitored system temperatures.

Each FEB uses commercial analog front-end (AFE) chips to amplify and digitize SiPM signals at 80~MS/s (Mega samples per second).
Data from the AFEs are processed by FPGAs and stored in buffer memory, awaiting triggers to initiate readout.
During data acquisition, triggers prompted the storage of pre- and post-trigger data for calibration and pedestal determination, with each event including 127~ADC samples per channel.
The data acquisition consisted of a two-phase ``cycle'': a 220 second live-time period for data collection, followed by a 20 second period for data transfer to the DAQ PC.
Typical event rates were 3~Hz for cosmic-ray triggers.

\section{Cosmic Ray Test Stand}\label{sec:teststand}

\begin{figure}
   \centering
   \includegraphics[width=0.75\linewidth]{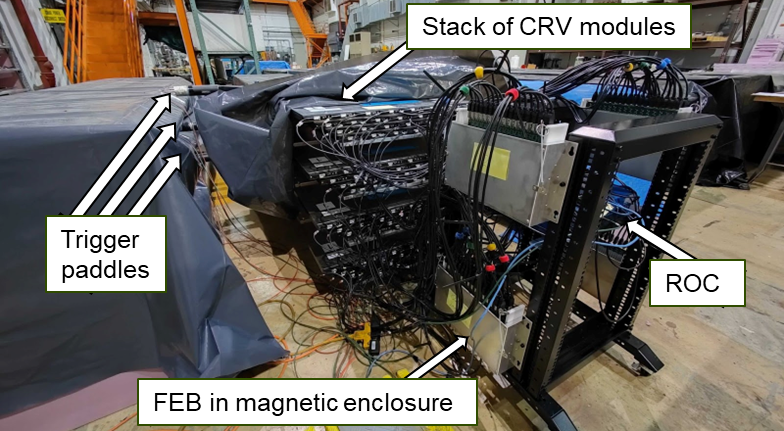}
     \label{subfig:Wideband_setup}
   \includegraphics[width=0.9\linewidth]{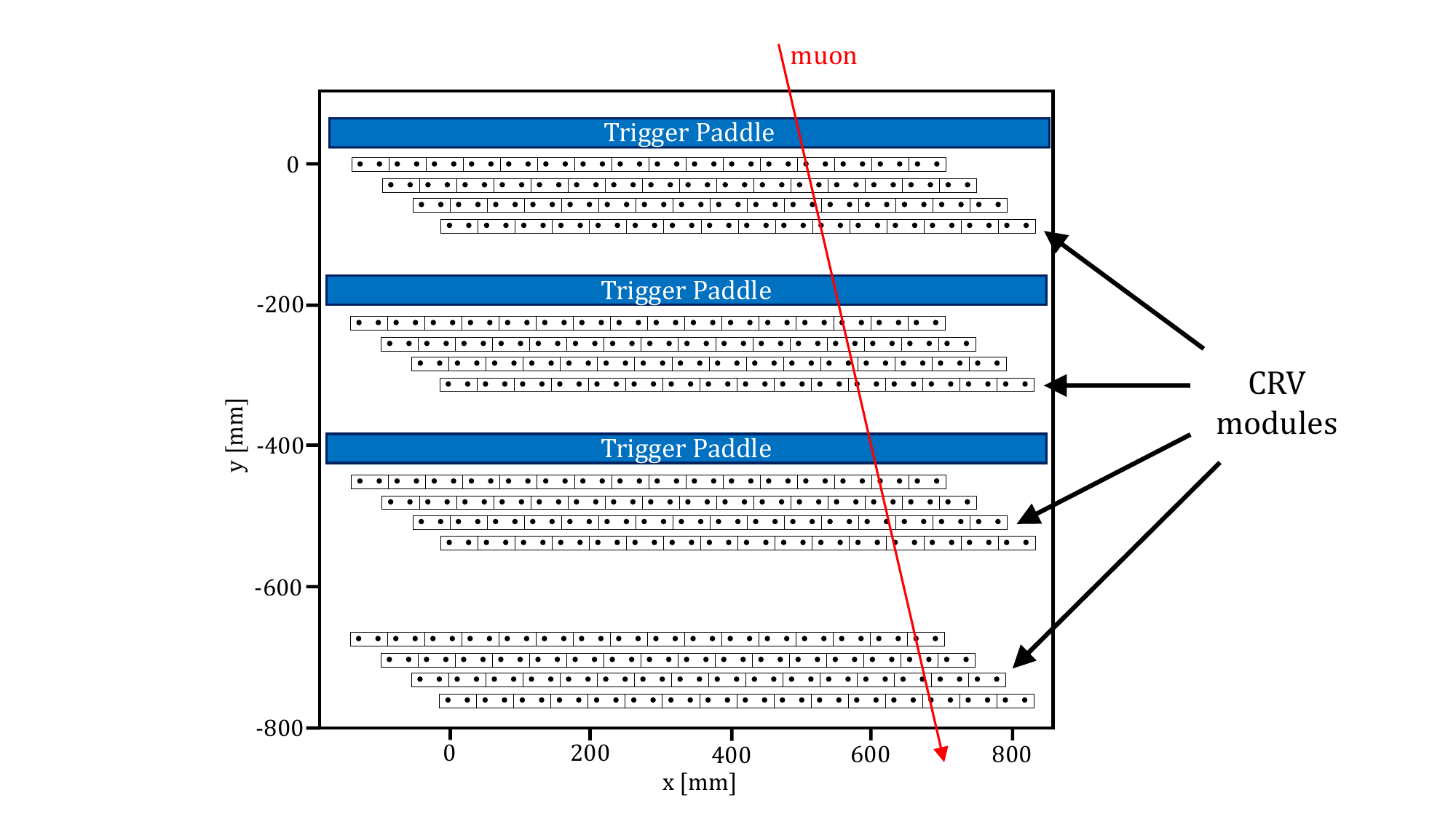}
     \label{subfig:Wideband_setup_cartoon}
     \caption{Top: the cosmic-ray test stand at Fermilab Wideband. Bottom: Cartoon of the cosmic-ray test stand end-view at Fermilab Wideband.}
   \label{fig:wideband}
 \end{figure}

At the Wideband hall at Fermilab, a subset of the modules are stored in stacks, as shown in Fig.~\ref{fig:wideband}.
Three $100\times10\times2$\,cm$^3$ plastic scintillator trigger paddles, read out with photomultiplier tubes\footnote{The scintillator paddles are EJ-200 plastic scintillator. A PMMA fishtail light guide connects the paddle to a Hamamatsu R329-02 photomultiplier tube. A 2\,mm thick EJ-560 optical interface pad is situated between the light guide and photomultiplier tube.},
were placed 1.0\,m from one of the module's readout ends, perpendicular to the direction of di-counters in the test modules.

In the studies presented here, three separate stacks of modules were used.
For the efficiency studies in Section~\ref{sec:efficiency} and the longitudinal position resolution studies in Section~\ref{sec:position}, we used a stack of four 6\,m long modules with 1.8\,mm diameter WLS fibers, which we henceforth refer to as CRV-T modules.
For the aging studies in Section~\ref{sec:aging}, we used a separate stack of four CRV-T modules, and a stack of four 4\,m long modules with 1.4\,mm diameter WLS fibers (referred to as CRV-L modules).
The modules used for our studies can be considered representative of all CRV production modules.

\begin{figure}
    \centering
    \includegraphics[width=0.8\linewidth]{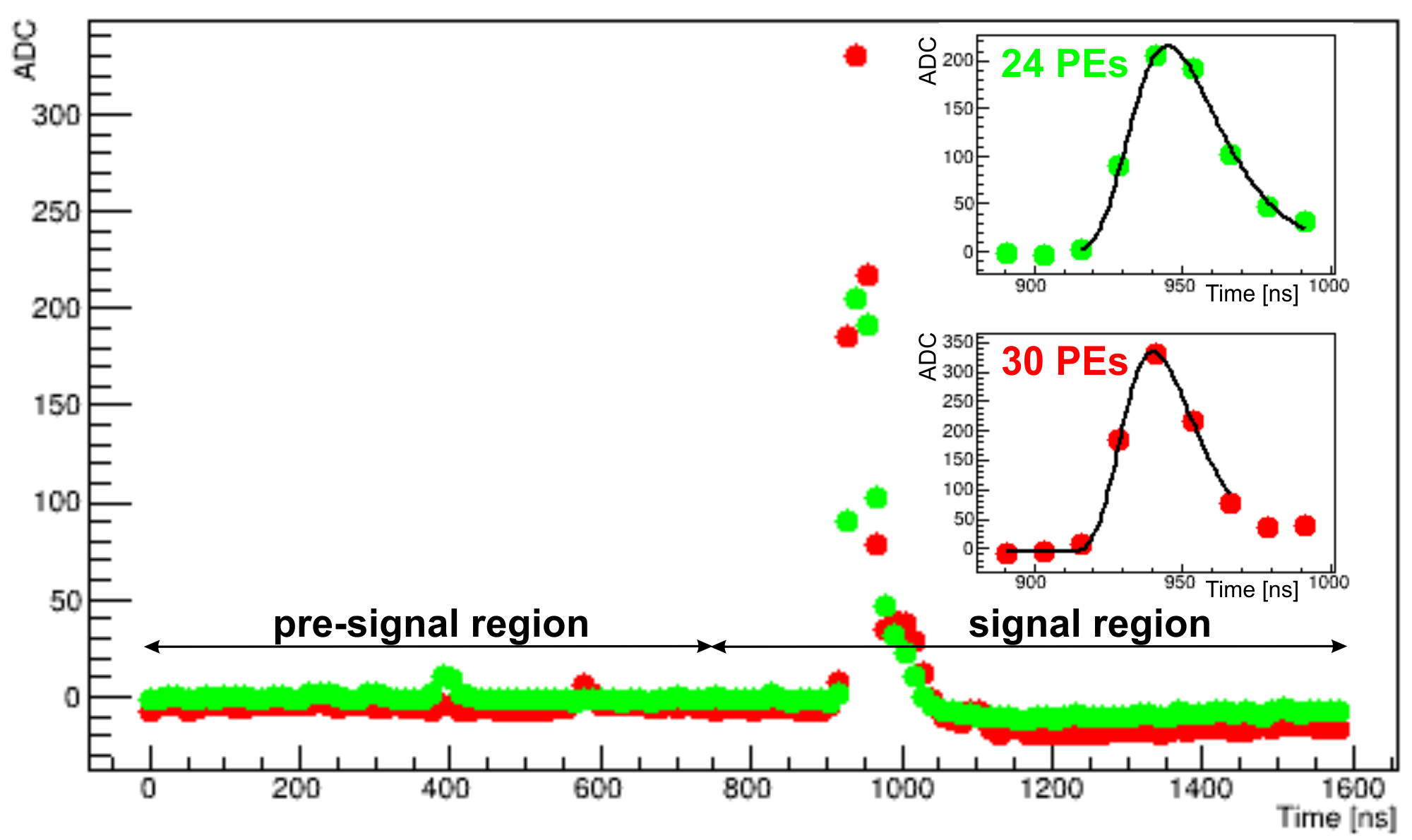}
\caption{Example waveform from two SiPM channels of a CRV counter, 127 ADC samples each. The pre-signal region shows dark-count pulses of 1\,PE in both channels at 400\,ns and 600\,ns. The signal region shows signal pulses with 24\,PEs and 30\,PEs in the respective channels. All pulses get fitted with a Gumbel function.}
\label{fig:waveform}
\end{figure}

To collect cosmic muon events in the cosmic-ray test stand, a three-fold coincidence among the trigger paddles was required within 100\,ns. 
The coincident trigger signals were distributed to the FEBs through the ROC.
Upon receiving a trigger, each SiPM channel read out a total of 127 samples at 80 MHz both before and after the trigger.
An example of the signal traces is shown in Fig.~\ref{fig:waveform}~\cite{crvsipm}.
Typically a trace contains 750~ns of data before the signal pulse (referred to as the pre-signal region), which allows for pedestal extraction.
In addition, the pre-signal region contains occasional pulses caused by SiPM dark noise, which is used for calibration purposes.
The data processing scheme is discussed in the following section.

\section{Data Processing}\label{sec:process}

The signals from the cosmic-ray test stand were calibrated and reconstructed. 
As the data were taken over several years with different temperatures, appropriate temperature corrections were applied. 
A detailed discussion of the calibration and reconstruction procedure, along with the determination of the temperature correction coefficients, can be found in Ref.~\cite{crvsipm}.
A summary of the procedure is given below.

The calibration procedure for the SiPM readout corrects for the temperature-dependent variations in gain and breakdown voltage.
The process involves measuring dark-noise spectra at the different temperatures and extracting key SiPM parameters such as gain, breakdown voltage, and photon detection efficiency.
Corrections are applied dynamically using real-time temperature readings to maintain precise charge reconstruction, crucial for reliable cosmic ray identification.

The reconstruction framework converts raw SiPM signals into meaningful physical observables, enabling efficient cosmic ray vetoing.
Signal processing includes pedestal subtraction, pulse shape fitting, and charge calibration. Following the pulse reconstruction, a linear track fit is done using the PE-weighted hit positions (which are the positions of the counters where the hits occurred). Hits with less than 5 PEs are excluded to avoid the impact of noise hits. 
The final reconstruction stage refines track parameters and assesses event quality, optimizing the veto’s efficiency in rejecting background particles while maintaining high signal fidelity.

\section{Efficiency Studies}\label{sec:efficiency}

In most areas covered by the CRV, a cosmic-ray muon is identified by a coincidence in space and time of at least three of the four layers of counters (3/4 efficiency).
The required efficiency for detecting such muons needs to be as high as 99.99\% in some critical regions of the CRV.
We describe below a measurement of the detection efficiency for cosmic-ray muons incident with a naturally occurring angular distribution.

\subsection{Event Selection}\label{subsec:event-sel}

To measure the efficiency to the required level, a very pure sample of cosmic-ray muons is needed.
In addition to the trigger requirement described in Section \ref{sec:teststand}, three of the four modules (called defining modules) in the stack were used in the offline analysis to define a good cosmic-ray muon
 used to measure the efficiency of the fourth module (the test module).
As in the real experiment, a cosmic-ray muon is considered detected in a layer of the module if the sum of PEs among the hits that are coincident in space and time on that layer exceeds a certain PE threshold. 
For the efficiency analysis, an additional linear fit of the hits in the four modules using the counter center positions weighted by the light yields was applied to assess whether a track was generated by a cosmic-ray muon.
Four selection cuts applied to data in the defining modules were used to define a good muon track:
\begin{enumerate}
    \item At least 10 PEs had to be deposited in each of the twelve layers of the defining modules,
    \item a linear fit to the hits in the defining modules had to have a $\chi^2/\text{n.d.f.} < 2.0$,
    \item there had to be less than $3000$ PEs in total over all the modules, and
    \item there had to be less than $40$ hits in total over all the modules.
\end{enumerate}
These criteria are based on the fact that the muons are minimum ionizing particles,
and hence they will leave linear tracks through the modules with an energy deposition proportional to their path length through the modules,
unlike, for example, events originating from electromagnetic or hadronic showers.
A total of $3.4\times10^6$ events passed our selection cuts from approximately two months of data collection.

\subsection{Measurement of the Track Stub Efficiency}\label{subsec:efficiency}

\begin{figure}[!htb]
    \centering
    \includegraphics[width=0.7\linewidth]{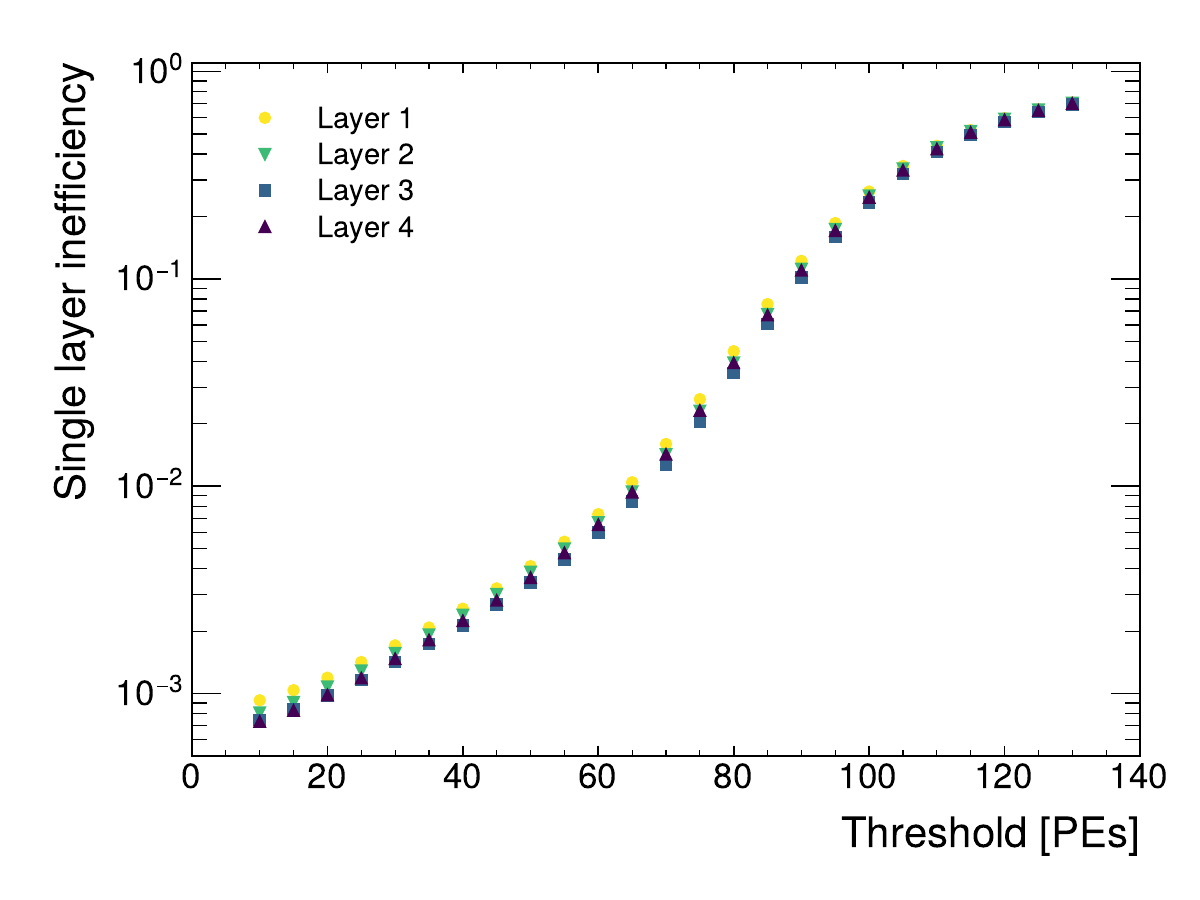}
    \caption{The single-layer inefficiency ($1-\varepsilon_{SL}$) of a CRV-T module as a function of the PE threshold.
             Note how each layer has a very similar efficiency.
    \label{fig:SingleLayerEfficiency}}
\end{figure}

After the event selection described above, the efficiency of a module was measured as a function of the PE threshold.
We examined two ways of determining the 3/4 efficiency of a module: 
an extrapolation using the single-layer efficiencies and a direct measurement of the 3/4 layer coincidence efficiency.
The single-layer inefficiency as a function of PE threshold is shown in Figure \ref{fig:SingleLayerEfficiency}.
The efficiencies of each layer are quite similar; hence we use their average value going forward.
Assuming each single-layer efficiency $\varepsilon_{SL}$ is independent, the efficiency $\varepsilon_{3/4}$ of at least 3 layers being hit is
\begin{equation}
   \varepsilon_{3/4} = \varepsilon_{SL}^4 + 4\times\varepsilon_{SL}^3(1-\varepsilon_{SL}) = 4\varepsilon_{SL}^3 - 3\varepsilon_{SL}^4. 
   \label{eq:eff}
\end{equation}

\begin{figure}[!htb]
    \centering
    \includegraphics[width=0.7\linewidth]{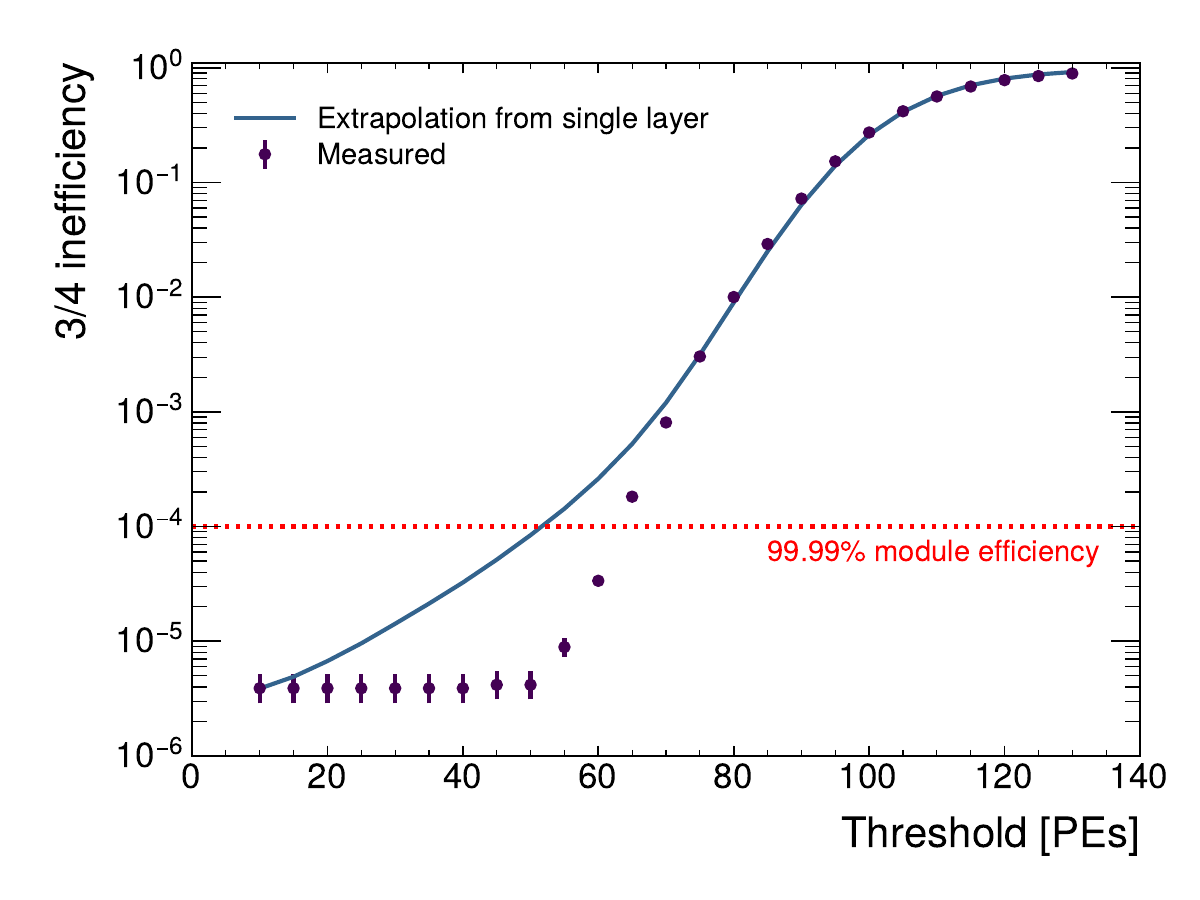}
    \caption{Inefficiency of hit coincidences in at least 3/4 layers of a module ($1-\varepsilon_{3/4}$) from direct measurements and 
    the extrapolation from single-layer efficiency using Eq.\,\ref{eq:eff}. For reference, the threshold when
    running the experiment is expected to be set at 10--15 PEs.
    \label{fig:ModuleEfficiency}}
\end{figure}

We compare this extrapolation from the single-layer efficiencies to the directly measured 3/4 efficiency in Figure~\ref{fig:ModuleEfficiency}.
They agree for high PE thresholds, but diverge significantly for lower PE thresholds, with the extrapolated 3/4 efficiency underestimating the efficiency.

\begin{figure}
    \centering
    \includegraphics[width=0.7\linewidth]{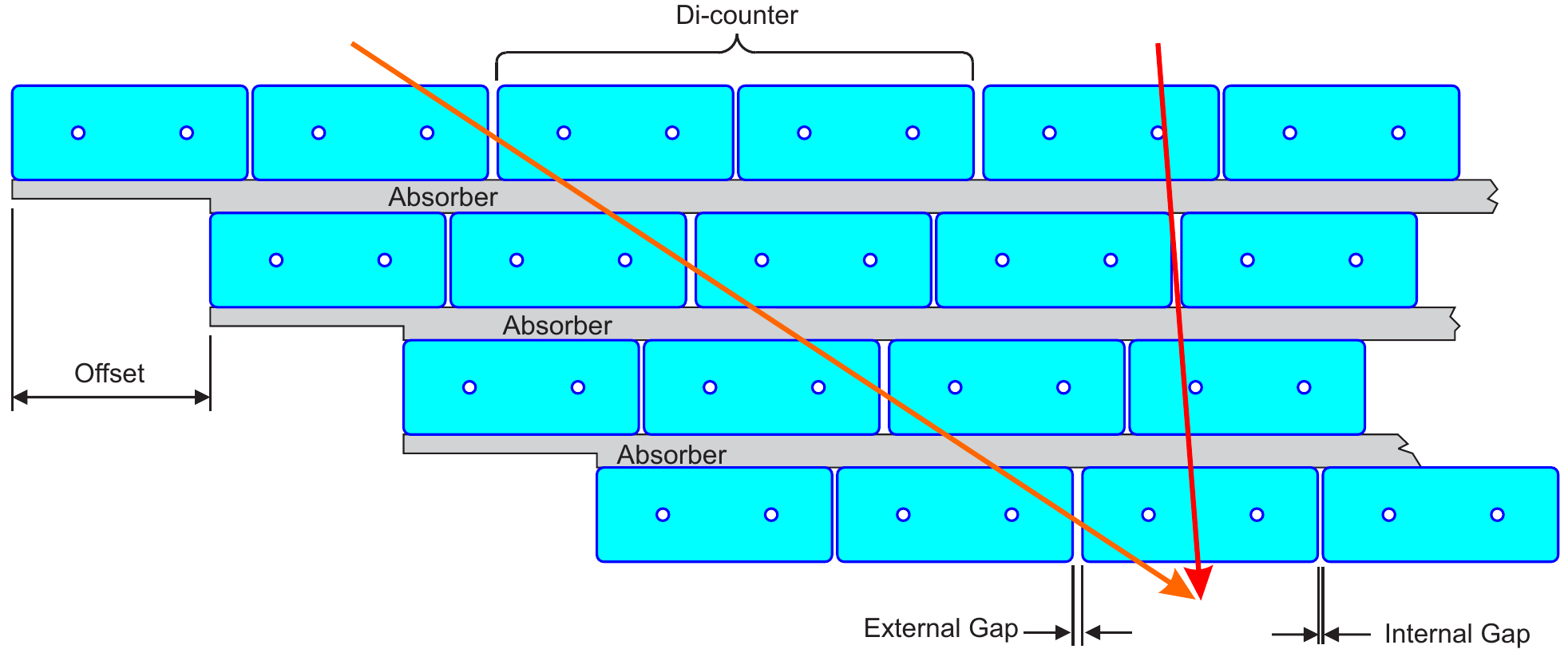}
    \caption{A cartoon illustrating two muon tracks passing through a module. The vertical red muon track goes through the external gap (exaggerated in width) 
    between two modules in the second layer.  However, due to the staggering of the layers, it is less likely to pass through another gap in a different layer.
    The orange muon track must be well inclined to pass through the external gap in all four layers, and hence goes through a significant amount of scintillator
    on the counters on either side of the gap.}
    \label{fig:gap_cartoon}
\end{figure}

The reason for the low PE threshold divergence is as follows.
The efficiency of a counter producing a hit is driven by the number of PEs produced in the counter, which is proportional to the path length of the muon through the counter.
At high thresholds, for a hit to be registered the muon has to pass through the counter and not a gap between counters --- even when traveling at an angle.
This inefficiency is layer-independent, and therefore the measured efficiency agrees well with the extrapolation.
However, as the PE threshold is lowered, it is much more likely that enough PEs will be produced to meet the threshold --- even for minimal path lengths on layers where the muon track goes through a gap --- so the inefficiency is primarily driven by muons going along the small gaps between di-counters.
Because of staggering in the module layers, a muon is much less likely to pass through a second gap if it has already passed through one, as illustrated in Fig.~\ref{fig:gap_cartoon}.\footnote{The module layer staggering of 42\,mm was determined by optimizing the 3/4 efficiency from extensive Monte Carlo studies of the response of top and side modules to cosmic-ray muons.}
The inefficiencies in the layers are no longer independent, as assumed in a simple extrapolation. 
As a result, the measured module efficiency exceeds the extrapolated value at such thresholds.
We observe that the measured 3/4 efficiency and the extrapolation start to diverge at a threshold of ${\sim}75$\,PEs, with the measured 3/4 efficiency being more efficient.
In the measured efficiency, we observe that the inefficiency plateaus below a 50 PE threshold.
This plateau is likely due to non-muon events that pass our selection cuts, which is supported by Monte Carlo studies, described in Section~\ref{sec:montecarlo}.

For reasonably and sufficiently low PE thresholds (10--50 PEs) both the measured efficiency and extrapolation from the single-layer efficiency exceeds the CRV's requirement of 99.99\%.
The operating PE threshold during Mu2e running will be 10--15 PEs, which is high enough to exclude noise-level hits and low enough to capture small energy depositions from cosmogenic particles.

\subsection{Monte Carlo Studies}\label{sec:montecarlo}

We used a Monte Carlo (MC) simulation to study the efficiency of our modules, in particular, to understand whether the inefficiency plateau at low PE thresholds observed in Fig.~\ref{fig:ModuleEfficiency} is due to muons or some other incident particle.

Generation of cosmic rays was performed using the CRY event generator~\cite{CRY2007}, which provides all cosmic particle production with the proper flux within a user-specified area and altitude.
The simulation of the CRV modules is based on the Geant4 library~\cite{geant4_2003,geant4_2006,geant4_2016}, which handles the particle transport through the modules and all particle interactions including the production of photons due to the scintillation process and Cherenkov effect in the scintillator counters.
It is also able to track these photons through the scintillator and the embedded wavelength-shifting fiber all the way to the SiPMs and simulate all processes such as wavelength shifting, refraction, and absorption.
However, tracking each individual photon through all counters requires an unreasonable high computation time, which is not feasible for simulations with a large number of events.
Therefore, lookup tables were produced (using Geant4) that store arrival probabilities and time-delay distributions for photons arriving at the SiPMs based on the location where the original particle went through the counter and the amount of energy that was deposited there.
Based on this information, a SiPM response with an appropriate pulse shape was simulated.
This step took into account the wavelength-dependent photon-detection efficiency of the SiPM pixels as well as the typical geometrical distribution of the photons at the end of the fiber.
The simulation of the SiPM response modeled the pixel discharge/recharge behavior, saturation effects, cross-talk, and dark counts.
The final SiPM pulse shape was the sum of the individual pulses of all SiPM pixels, where the single-pixel pulse shape was determined by an independent simulation based on the properties of the SiPM.
The last step of the simulation used the responses of the SiPMs of all counters and generated a digitized waveform similar to the waveforms produced by the experiment.
The simulation was tuned to match the results of test-beam studies and studies with cosmic rays~\cite{Mu2e:2017lae}.

\begin{figure}[!htb]
    \centering
    \includegraphics[width=0.7\linewidth]{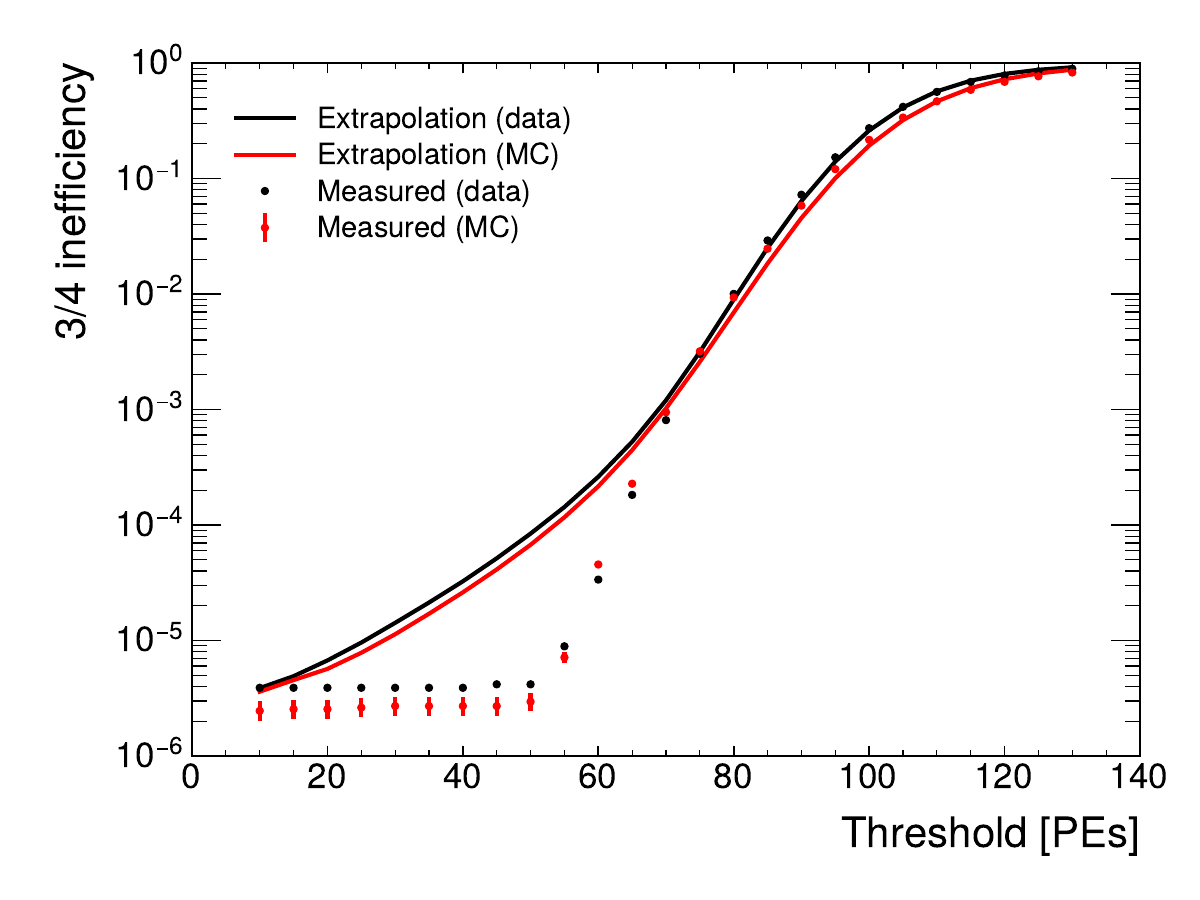}
  \caption{The directly measured 3/4 inefficiency and the extrapolated 3/4 inefficiency from the single-layer efficiency for MC and data.}
  \label{fig:EffMCvsData}
\end{figure}

We generated $5.6\times10^7$ simulated events and passed them through the selection cuts described in Section~\ref{subsec:event-sel}.
The measured 3/4 efficiency and extrapolation from single-layer efficiency of the MC sample, with a comparison to data, is shown in Figure~\ref{fig:EffMCvsData}.
The simulation well reproduces the module efficiency we observe in data, including the plateau at low PE thresholds.

The simulation reproduces the plateau in efficiency observed in data, although there are still discrepancies between the efficiencies measured in simulation and data.
Sources of discrepancy may come from how the experimental environment is being modeled in simulation, which was found to be an important factor in matching the results to data.
For technical reasons, the simulation places the module stacks inside of the Mu2e experimental hall, which was modified to more closely mimic the Wideband facility.
Differences in the geometry and material structure of the environment between simulation and data may contribute to differences in a way that is not easily quantifiable as a systematic uncertainty.
Another potential source of error in the environmental modeling is discrepancies between the distribution of cosmic-ray events in simulation and reality.
We continue to improve the accuracy of our simulation, particularly for its use in simulating the Mu2e experimental configuration.
But for the purpose of this paper, it is not critical to perfectly model the module efficiency in simulation;
instead, the reproduction of the plateau in efficiency at low PE thresholds is the essential component of our simulation studies.
\begin{figure}[!htb]
    \centering
    \includegraphics[width=0.53\linewidth]{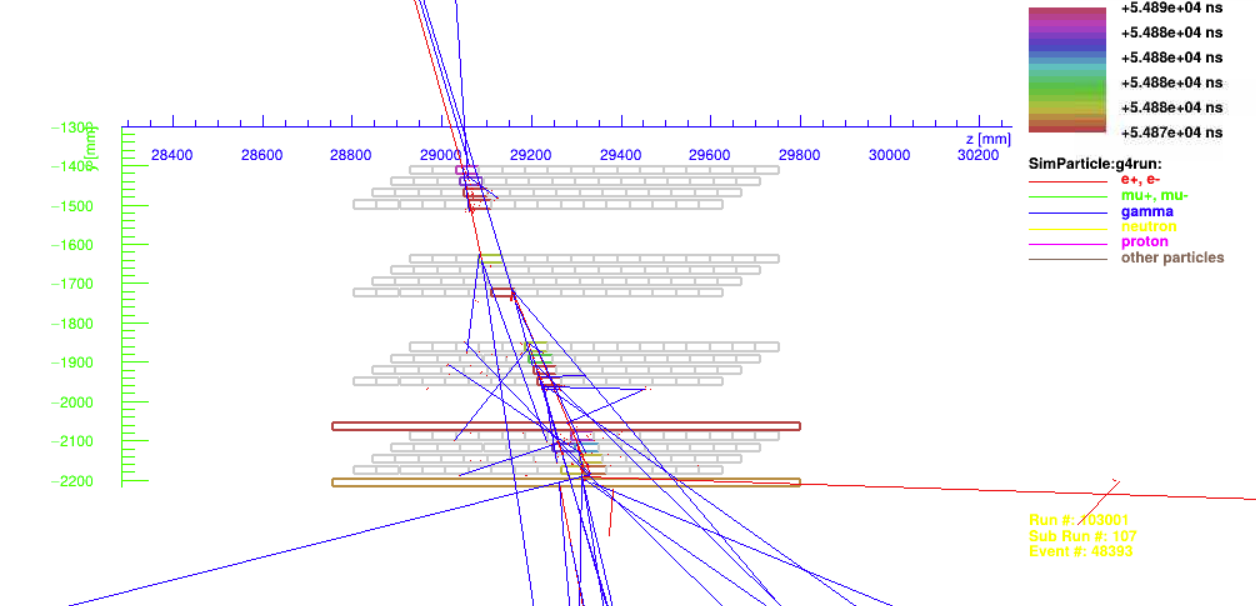}
    \includegraphics[width=0.45\linewidth]{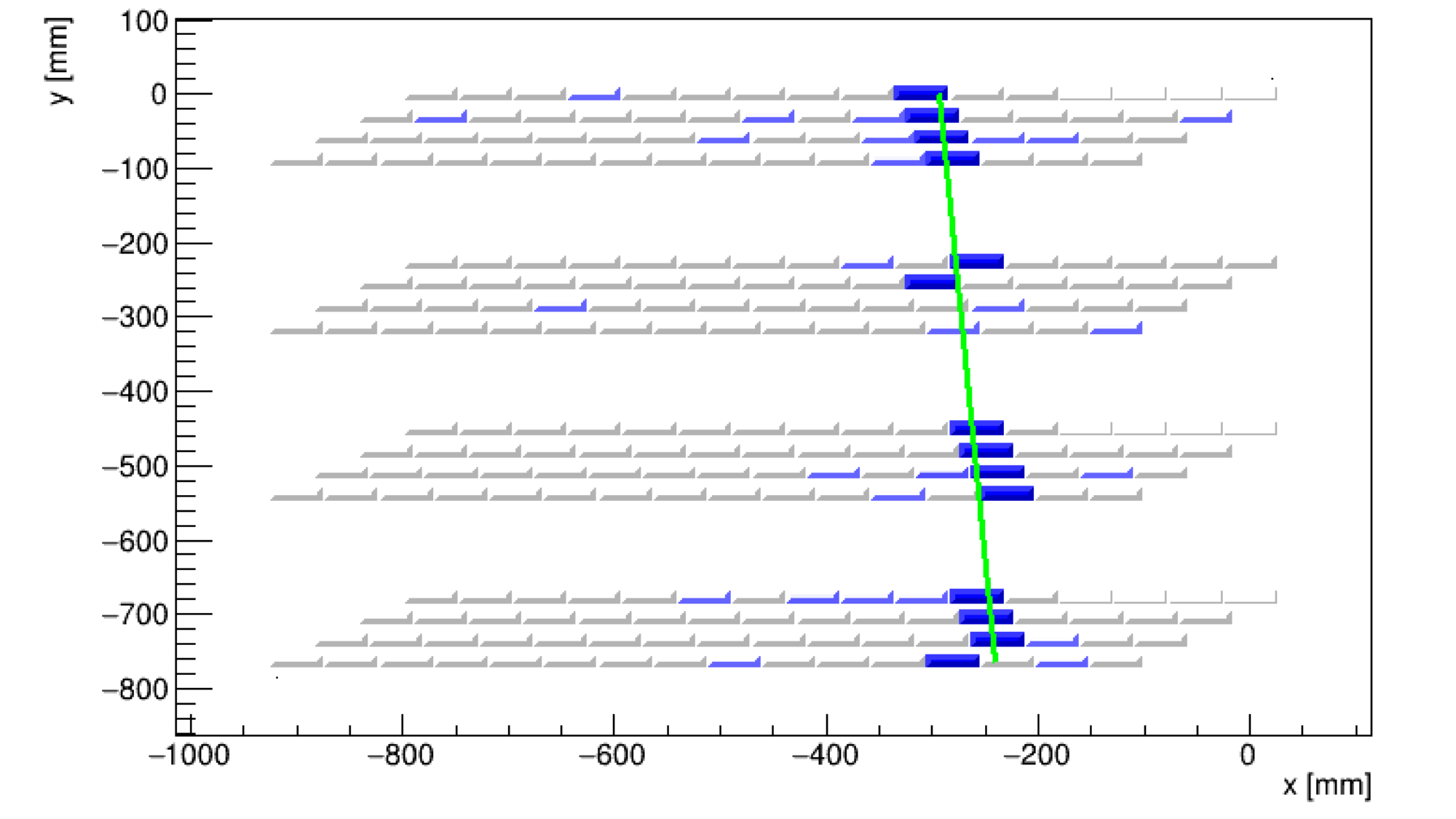}
  \caption{Example event displays for a simulated event (left) with similar hit patterns to a data event (right). These events pass selection cuts but do not pass the 3/4 layer veto requirement for low PE thresholds.}
  \label{fig:EventDisplays}
\end{figure}

Upon inspection of events in the plateau of the measured MC efficiency, we find that all of the inefficient events originate from incident particles that are not muons.
Figure~\ref{fig:EventDisplays} shows an example of an inefficient simulated event at low PE thresholds with hit patterns similar to an inefficient event from data.
The example event, and all inefficient events in our simulation, originate from incident particles other than muons.
Although events such as these arise in our test stand, they are unlikely to be a source of inefficiency for the CRV during Mu2e operations due to concrete shielding blocks above the experimental apparatus.
The suppression of these events from shielding blocks is supported by further simulation studies with concrete slabs situated above the modules.

\section{Aging of Mu2e CRV Modules}\label{sec:aging}
It has long been established that the light yield from plastic scintillators decreases over time~\cite{barnaby1962}, a phenomenon known as scintillator aging.
Since the light yield of the CRV directly affects veto efficiency, it is crucial to accurately characterize the aging rate of Mu2e CRV modules. Ensuring the CRV system maintains the required veto efficiency throughout the experiment's lifetime depends on understanding this aging process. During the design phase, an aging rate of approximately 3\% per year was anticipated, based on observations from the MINOS experiment~\cite{minos2008}. Initial studies using prototype counters showed much higher aging rates. Consequently, Mu2e CRV aging studies commenced promptly once production scintillation counters became available.

It is important to note that plastic scintillators age non-uniformly, as demonstrated by findings from the T2K experiment~\cite{t2k2022}. Typically, scintillators undergo a rapid aging phase during the first 1–2 years post-fabrication, after which the rate of aging stabilizes to a slower, long-term trend. Our goal was to obtain a definitive measurement of this long-term aging rate for the CRV modules. To achieve this, we established a series of cosmic-ray test stands at the Fermilab Wideband facility. Given that the modules were fabricated from scintillator material extruded in 2018 and the initial test stand was installed in 2021, our measurements are expected to reflect only the long-term aging behavior, excluding the initial rapid aging phase.

The configuration of the cosmic-ray test stands used for aging studies is described in Section~\ref{sec:teststand}. The first test stand, deployed in July 2021, consists of four CRV-L modules equipped with 1.4\,mm diameter wavelength-shifting (WLS) fibers. Initially, only two readout FEBs were available, allowing readout from a single end of one module. 
Light-yield measurements from these channels have been continuously recorded over nearly three years. In October 2022, four additional FEBs were installed to expand the readout coverage.

A second test stand, featuring CRV-T modules with 1.8mm diameter WLS fibers, was commissioned in August 2022. This setup includes six readout FEBs and provided additional data to evaluate the aging characteristics across different module configurations.

\begin{figure}
\centering
\includegraphics[width=0.7\textwidth]{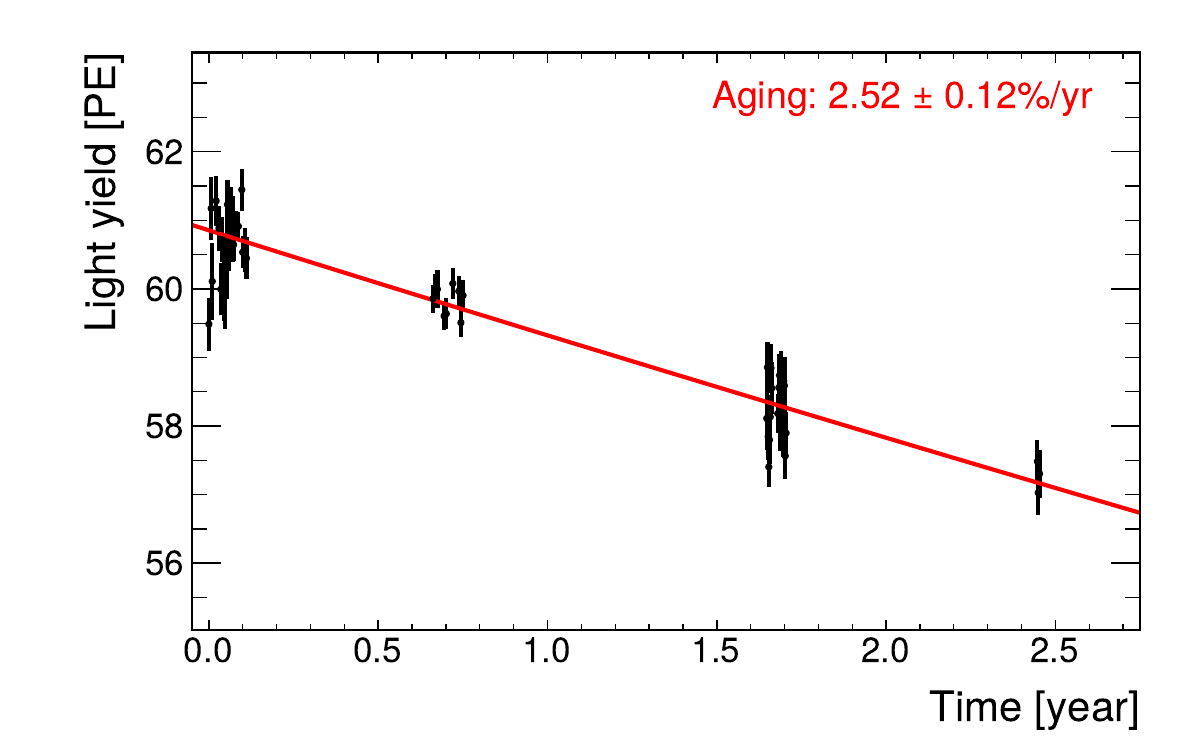}
\caption{Light yield over time of an example channel from a CRV-T module.} \label{fig:aging_example}
\end{figure}

The data collected from the cosmic-ray test stands were calibrated using the procedure outlined in Section~\ref{sec:process}. Photoelectron (PE) yields were calculated for each run over time and used to extract aging rates, as illustrated in Fig.\ref{fig:aging_example}. For each channel (SiPM), the light-yield data were fitted with the following function:
\begin{equation}
A_{0} (1 - r/100)^{t},
\end{equation}
where $A_{0}$ is the initial light yield in PEs at time $t = 0$, $r$ is the aging rate in percent per year, and $t$ is the elapsed time in years since the first measurement.

\begin{figure}
\centering
\includegraphics[width=0.7\textwidth]{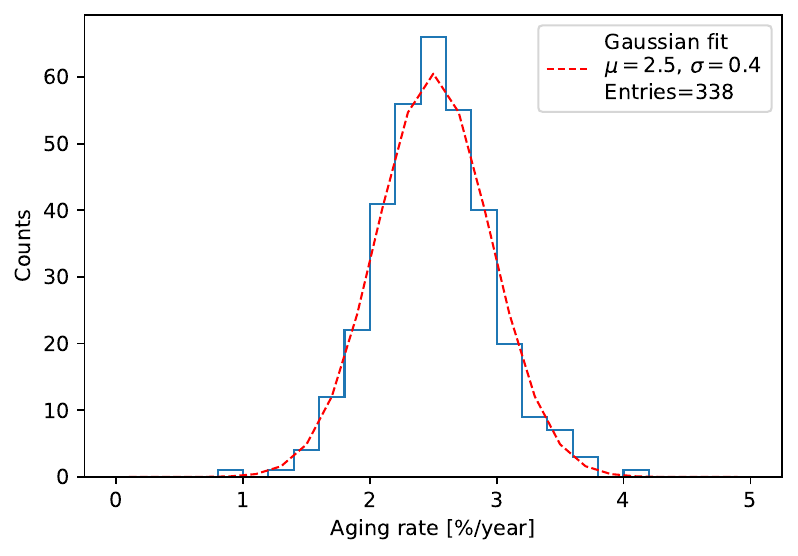}
\caption{Aging rate of CRV-T modules using cosmic-ray muons incident one meter from the readout.} \label{fig:aging_distribution}
\end{figure}

We observed an average aging rate of ($2.5\pm0.4$)\%/yr over $\sim$2.5 years from a measurement of 338 channels from CRV-T modules, as shown in Figure~\ref{fig:aging_distribution}.
A measurement on a set of 42 counters from a CRV-L module showed an aging rate of ($3.0\pm1.1$)\%/yr over a $\sim$3-year period.

The scintillator counters included in this study come from different batches, which we observe to have slightly different aging rates.
When we measure the aging rate of individual modules in the configuration, we observe normally distributed aging rates.
The $\pm0.4$\%/yr variance we observe in the aging distribution encompasses some of this systematic variation in the scintillator counter batches.

\section{Extrapolation to Overall CRV Efficiency}\label{sec:extrapolate}

The expected background for Mu2e Run I from cosmic-ray-induced backgrounds was previously determined to be $0.046\,\pm\,0.010$ events, which assumed the aging rate of 8.7\%/yr that had been measured from pre-production counters~\cite{SU2020}.
Keeping the expected background below one event for the entirety of the Mu2e runs requires a CRV efficiency of 99.99\%.
We find that the efficiency of CRV modules should exceed 99.99\% for sufficiently (but reasonably) low PE thresholds and for an aging rate of modules of $\sim3$\%/yr, which is significantly lower than earlier measurements.
Based on these results, we expect the CRV to maintain an efficiency that keeps the expected cosmic-ray-induced background below one event over Mu2e's anticipated run time. A systematic evaluation of the overall CRV efficiency along with an updated evaluation of cosmic-ray-induced backgrounds will be presented in a forthcoming publication. 

\section{Longitudinal Position Resolution Along a Counter}\label{sec:position}
The longitudinal position along the length of a scintillator counter can be
determined by the time of arrival (TOA) of the scintillation light to the SiPMs on each
end of the counter. To determine the position from the TOA information, a modified
version of the cosmic-ray test stand was used. Eight FEB boards were used in the study, 
but instead of reading out the modules on one end, two 6 m long modules were read out on both ends. 
The small timing differences between the FEBs were measured and corrected for.
Trigger paddles placed 1.0\,m from one end of the modules were used for event triggering.
As the trigger paddles are 10\,cm wide, the expected position of cosmic-ray muons from one 
end is $105\pm5$\,cm. 

\begin{figure}
\centering
\includegraphics[width=0.7\textwidth]{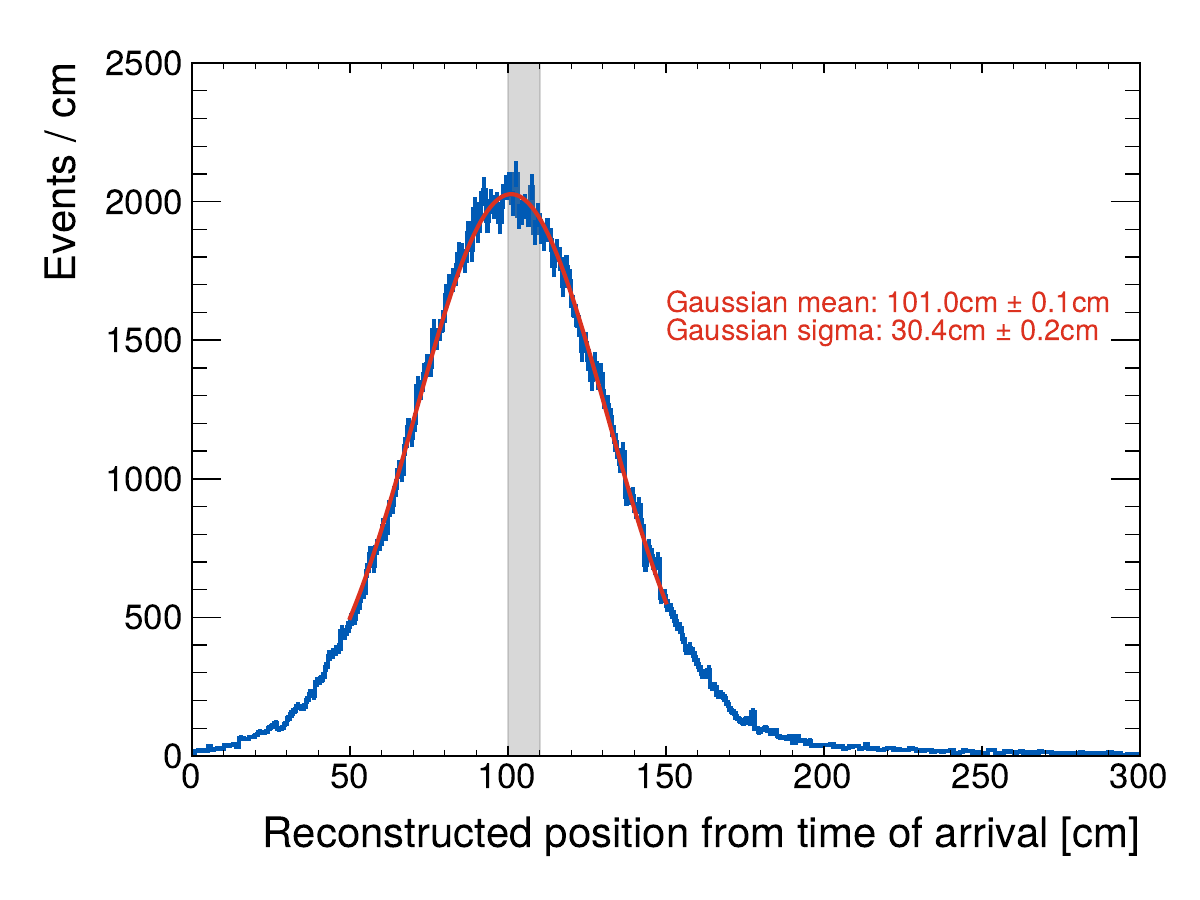}
\caption{The reconstructed position of cosmic-ray events from the end closest
to the trigger paddles using the light arrival time differences from the two ends of a
module and assuming a speed of light in the WLS fiber of 17.3\,cm/ns. Events are
triggered from a coincidence in two trigger paddles, represented by the shaded gray
region.} 
\label{fig:position_resolution}
\end{figure}

The position from end-1, $x_1$, of a particle passing through a counter is determined from the light
arrival time difference, $\Delta{t_{12}} = t_1 - t_2$, between end-1 and end-2 using:
\begin{equation}
x_1 = \frac{1}{2}(L + c_f\Delta{t_{12}}),
\end{equation}
where $L$ is the counter length and $c_f$ is the speed of light in the fiber.
The speed of light in the WLS fiber was determined to be 17.3\,cm/ns from test-beam
studies on counters with a 1.4\,mm diameter WLS fiber \cite{Mu2e:2017lae}. 
Figure~\ref{fig:position_resolution} shows
the reconstructed position based on the light arrival time difference. A gaussian 
distribution fitted to the distribution was used to determine the mean and
standard deviation. The mean of 101.0\,cm is consistent with
the position of the trigger paddles at $105\pm5$\,cm from the near end of the module.
The uncertainty of the measurement is 30.4\,cm, which is about twice as large
as the uncertainty from similar measurements using a test beam \cite{Mu2e:2017lae}. 
The 10\,cm width of the trigger paddles contributes to this larger uncertainty,
as well as the much shorter, pre-production counters placed in the test beam.

\section{Summary}\label{sec:summary}
Measurements of the cosmic-ray muon detection efficiency and module aging rate for the Mu2e cosmic ray veto detector (CRV) have been reported. Using test stands consisting of production Mu2e CRV modules, we confirmed that, at appropriate photoelectron (PE) threshold levels, the efficiency of CRV modules can reach the designed efficiency requirement of 99.99\%. Monte Carlo studies show good agreement with the measurements regarding the relationship between module detection efficiency and the PE threshold, explaining features in this relationship caused by module geometry and non-muonic components in the cosmic rays. The long-term aging rates of the production modules, with counters embedded with wavelength-shifting (WLS) fibers of two different diameters, were measured over multiple years. A similar aging rate of approximately 2.5--3\% per year was reported for both types of counters. With the measured detection efficiency and aging rate, we expect that the CRV will remain over 99.99\% efficient and will be capable of suppressing the cosmic-ray-induced background for the Mu2e experiment to below one event over Mu2e's anticipated run time. In addition, the longitudinal position resolution along a counter was measured to be 30.4\,cm using a modified version of the test stand.

\section{Acknowledgments}
We are grateful for the vital contributions of the Fermilab staff and the technical staff of the participating institutions. This work was supported by the US Department of Energy; the Istituto Nazionale di Fisica Nucleare, Italy; the Science and Technology Facilities Council, UK; the Ministry of Education and Science, Russian Federation; the National Science Foundation, USA; the National Science Foundation, China; the Helmholtz Association, Germany; and the EU Horizon 2020 Research and Innovation Program under the Marie Sklodowska-Curie Grant Agreement Nos.  734303, 822185, 858199, 101003460, and 101006726. This document was prepared by members of the Mu2e Collaboration using the resources of the Fermi National Accelerator Laboratory (Fermilab), a U.S. Department of Energy, Office of Science, HEP User Facility. Fermilab is managed by Fermi Forward Discovery Group, LLC, acting under Contract No. 89243024CSC000002.
This material is based upon work supported by the U.S. Department of Energy, Office of Science, Office of Workforce Development for Teachers and Scientists, Office of Science Graduate Student Research (SCGSR) program. The SCGSR program is administered by the Oak Ridge Institute for Science and Education for the DOE under contract number DE‐SC0014664. 

\bibliographystyle{elsarticle-num}
\bibliography{references} 

\begin{thebibliography}{10}
\expandafter\ifx\csname url\endcsname\relax
  \def\url#1{\texttt{#1}}\fi
\expandafter\ifx\csname urlprefix\endcsname\relax\def\urlprefix{URL }\fi
\expandafter\ifx\csname href\endcsname\relax
  \def\href#1#2{#2} \def\path#1{#1}\fi

\bibitem{Mu2e:2014fns}
L.~Bartoszek, et~al., {Mu2e Technical Design Report}, arXiv:1501.05241 [physics.ins-det] (2014).
\newblock \href {http://arxiv.org/abs/1501.05241} {\path{arXiv:1501.05241}}.

\bibitem{Marciano:2008zz}
W.~J. Marciano, T.~Mori, J.~M. Roney, {Charged Lepton Flavor Violation Experiments}, Ann. Rev. Nucl. Part. Sci. 58 (2008) 315--341.
\newblock \href {https://doi.org/10.1146/annurev.nucl.58.110707.171126} {\path{doi:10.1146/annurev.nucl.58.110707.171126}}.

\bibitem{SU2020}
F.~Abdi, et~al., {Mu2e Run I Sensitivity Projections for the Neutrinoless $\mu^- \to e^-$ Conversion Search in Aluminum}, Universe 9~(1) (2023) 54.
\newblock \href {http://arxiv.org/abs/2210.11380} {\path{arXiv:2210.11380}}, \href {https://doi.org/10.3390/universe9010054} {\path{doi:10.3390/universe9010054}}.

\bibitem{Boi:2021yxw}
S.~Boi, {Design and Fabrication of a Novel Large-Area, High-Efficiency Cosmic Ray Veto Detector for the Mu2e Experiment}, FERMILAB-THESIS-2021-29 (2021).
\newblock \href {https://doi.org/10.18130/0nf5-sw49} {\path{doi:10.18130/0nf5-sw49}}.

\bibitem{crvsipm}
L.~Curtis, et~al., {Temperature-Dependent Calibration Procedures for the Silicon Photomultiplier Readout of the Cosmic Ray Veto Detector for the Mu2e Experiment}, Nucl. Instrum. Meth. A 1081 (2026) 170809.
\newblock \href {https://doi.org/https://doi.org/10.1016/j.nima.2025.170809} {\path{doi:https://doi.org/10.1016/j.nima.2025.170809}}.

\bibitem{CRY2007}
C.~Hagmann, D.~Lange, D.~Wright, {Cosmic-Ray Shower Generator ({CRY}) for Monte Carlo Transport Codes}, in: 2007 IEEE Nuclear Science Symposium Conference Record, Vol.~2, 2007, pp. 1143--1146.
\newblock \href {https://doi.org/10.1109/NSSMIC.2007.4437209} {\path{doi:10.1109/NSSMIC.2007.4437209}}.

\bibitem{geant4_2003}
S.~Agostinelli, et~al., {Geant4—a Simulation Toolkit}, Nucl. Instrum. Meth. A 506~(3) (2003) 250--303.
\newblock \href {https://doi.org/https://doi.org/10.1016/S0168-9002(03)01368-8} {\path{doi:https://doi.org/10.1016/S0168-9002(03)01368-8}}.

\bibitem{geant4_2006}
J.~Allison, et~al., {Geant4 Developments and Applications}, IEEE Trans. Nucl. Sci. 53~(1) (2006) 270--278.
\newblock \href {https://doi.org/10.1109/TNS.2006.869826} {\path{doi:10.1109/TNS.2006.869826}}.

\bibitem{geant4_2016}
J.~Allison, et~al., {Recent Developments in Geant4}, Nucl. Instrum. Meth. A 835 (2016) 186--225.
\newblock \href {https://doi.org/https://doi.org/10.1016/j.nima.2016.06.125} {\path{doi:https://doi.org/10.1016/j.nima.2016.06.125}}.

\bibitem{Mu2e:2017lae}
A.~Artikov, et~al., {Photoelectron Yields of Scintillation Counters with Embedded Wavelength-Shifting Fibers Read Out With Silicon Photomultipliers}, Nucl. Instrum. Meth. A 890 (2018) 84--95.
\newblock \href {http://arxiv.org/abs/1709.06587} {\path{arXiv:1709.06587}}, \href {https://doi.org/10.1016/j.nima.2018.02.023} {\path{doi:10.1016/j.nima.2018.02.023}}.

\bibitem{barnaby1962}
C.~F. Barnaby, J.~C. Barton, {Ageing of Plastic Scintillators}, J. Sci. Instrum. 39~(4) (1962) 176.
\newblock \href {https://doi.org/10.1088/0950-7671/39/4/426} {\path{doi:10.1088/0950-7671/39/4/426}}.

\bibitem{minos2008}
D.~Michael, et~al., {The Magnetized Steel and Scintillator Calorimeters of the MINOS Experiment}, Nucl. Instrum. Meth. A 596~(2) (2008) 190--228.
\newblock \href {https://doi.org/https://doi.org/10.1016/j.nima.2008.08.003} {\path{doi:https://doi.org/10.1016/j.nima.2008.08.003}}.

\bibitem{t2k2022}
K.~Abe, et~al., {Scintillator Ageing of the {T2K} Near Detectors from 2010 to 2021}, J. Instrum. 17~(10) (2022) P10028.
\newblock \href {https://doi.org/10.1088/1748-0221/17/10/P10028} {\path{doi:10.1088/1748-0221/17/10/P10028}}.

\end{thebibliography}

\newpage
\pagenumbering{arabic}

\end{document}